\documentclass[epj]{svjour}
\usepackage{latexsym}
\usepackage{graphics}
\usepackage{graphicx}

\begin{document}
\title{Parity Violation, the Neutron Radius of Lead, and Neutron Stars}
\author{J. Piekarewicz}
\institute{Department of Physics, Florida State
           University, Tallahassee, FL 32306, USA}
\date{Received: date / Revised version: date}
%
\abstract{The neutron radius of a heavy nucleus is a fundamental
nuclear-structure observable that remains elusive. Progress in this
arena has been limited by the exclusive use of hadronic probes that 
are hindered by large and controversial uncertainties in the reaction 
mechanism. The Parity Radius Experiment at the Jefferson Laboratory 
offers an attractive electro-weak alternative to the hadronic program 
and promises to measure the neutron radius of $^{208}$Pb accurately 
and model independently via parity-violating electron scattering. In 
this contribution we examine the far-reaching implications that such 
a determination will have in areas as diverse as nuclear structure,
atomic parity violation, and astrophysics.
} 
\maketitle
\section{Introduction}
\label{introduction}

The nucleus of ${}^{208}$Pb is 18 order of magnitudes smaller and 55
orders of magnitude lighter than a neutron star. Yet remarkably, both
the neutron radius of ${}^{208}$Pb as well as the radius of a neutron
star depend critically on our (incomplete) knowledge of the equation 
of state of neutron-rich matter. The emergence of such a correlation 
among objects of such a disparate size is not difficult to understand. 
Heavy nuclei develop a neutron-rich skin as a result of a large neutron 
excess and a large Coulomb barrier that reduces the proton density at 
the surface of the nucleus. Thus the thickness of the neutron skin 
depends on the pressure that pushes neutrons out against surface 
tension~\cite{Brown:2000}. It is this same pressure that supports a 
neutron star against gravitational 
collapse~\cite{Lattimer:2000nx,Steiner:2004fi}.  Thus models with
thicker neutron skins often produce neutron stars with larger
radii~\cite{Horowitz:2001ya}.

Attempts at mapping the neutron distribution of ${}^{208}$Pb, or of
any other heavy nucleus, using hadronic probes have been met with
limited success. Although highly mature and successful, the hadronic
program will never attain the precision status that the electro-weak
program enjoys. This is due to the large and controversial
uncertainties in the reaction
mechanism~\cite{Ray:1985yg,Ray:1992fj}. A particularly illustrative
example of this situation is provided by the proton and neutron radii
of ${}^{208}$Pb. While elastic electron scattering experiments have
determined the charge radius of ${}^{208}$Pb to better than
0.001~fm~\cite{Fricke:1995}, realistic estimates place the uncertainty
in the neutron radius at about 0.2~fm~\cite{Horowitz:1999fk}.

The enormously successful parity violating program at the Jefferson
Laboratory~\cite{Aniol:2005zf,Aniol:2005zg} provides an attractive
electro-weak alternative to the hadronic program. Indeed, the Parity
Radius Experiment (PREX) at the Jefferson Laboratory aims to measure
the neutron radius of $^{208}$Pb accurately (to within $0.05$~fm) and
model independently via parity-violating electron
scattering~\cite{Horowitz:1999fk}. Parity violation at low momentum
transfers is particularly sensitive to the neutron density because the
$Z^0$ boson couples primarily to neutrons. Moreover, the
parity-violating asymmetry, while small, can be interpreted with as
much confidence as conventional electromagnetic scattering
experiments. PREX will provide a unique observational constraint on
the thickness of the neutron skin of a heavy nucleus. We note that
since first proposed in 1999, many of the technical difficulties
intrinsic to such a challenging experiment have been
met~\cite{Michaels:2005}. For further details on the status of the
experiment, see the contribution from Robert Michaels to these
proceedings.

\section{Formalism}
\label{formalism}

The starting point for the calculation of both the properties of finite
nuclei, their self-consistent linear response, and the structure and
dynamics of neutron stars is based on a relativistic density functional.
The underlying Lagrangian density includes an isodoublet nucleon field 
interacting via the exchange of two isoscalar mesons --- a scalar and a 
vector --- one isovector meson, and the photon. Details of this model
may be found in Refs.~\cite{Serot:1984ey,Serot:1997xg}. In addition to 
meson-nucleon interactions, the Lagrangian density must be supplemented 
by nonlinear meson interactions that are responsible for a softening of
the equation of state of symmetric nuclear matter at both normal and
high densities~\cite{Mueller:1996pm}. Of particular relevance to the
present contribution is an effective coupling constant (denoted by
$\Lambda_{\rm v}$) that induces isoscalar-isovector mixing and has 
been added to tune the poorly-known density dependence of the symmetry
energy~\cite{Horowitz:2001ya,Horowitz:2000xj}. As a result of the
strong correlation between the neutron radius of heavy nuclei and the
pressure of neutron-rich matter~\cite{Brown:2000,Furnstahl:2001un},
the neutron skin of a heavy nucleus is also highly sensitive to changes 
in $\Lambda_{\rm v}$. Hence, the interacting Lagrangian density of 
Ref.~\cite{Mueller:1996pm} has been supplemented by the following 
term: 
\begin{equation}
{\cal L}_{\rm int} =
    \Lambda_{\rm v}
    \Big(g_{\rho}^{2}\,{\bf b}_{\mu}\cdot{\bf b}^{\mu}\Big)
    \Big(g_{\rm v}^2V_{\mu}V^\mu\Big) \;,
\label{Lagrangian}
\end{equation}
where $V^{\mu}$ and ${\bf b}^{\mu}$ denote isoscalar and
isovector vector fields, respectively.

\section{Results}
\label{results}

\begin{figure}
\vspace{0.50in}
\includegraphics[width=3.25in,angle=0]{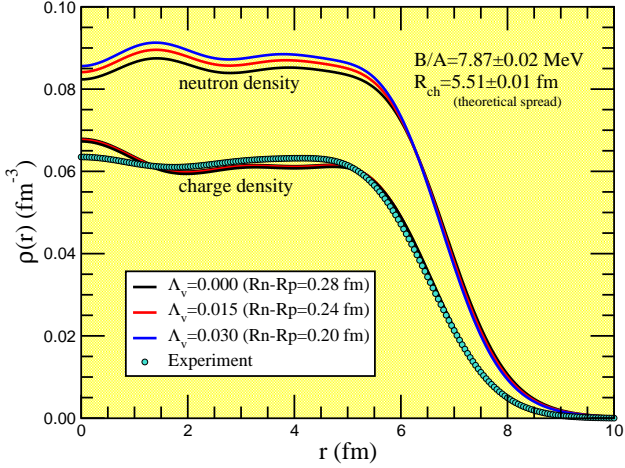}
\caption{Proton (charge) and neutron (point) densities for ${}^{208}$Pb 
 using a variety of values for the isoscalar-isovector coupling constant 
 $\Lambda_{\rm v}$.}
\label{Fig1}       
\end{figure}

We start this section by displaying in Fig.~\ref{Fig1} the impact of
the addition of the empirical coupling constant $\Lambda_{\rm v}$ on
both the charge and point-neutron densities of ${}^{208}$Pb. An
important constraint that must be satisfied by the addition of any new
coupling constant into the Lagrangian is that the success of the
effective model in reproducing well determined ground-state
observables is maintained. Figure~\ref{Fig1} indicates that this is
indeed the case for the charge density --- an observable largely
insensitive to the value of $\Lambda_{\rm v}$. In contrast, the
changes in $\Lambda_{\rm v}$ depicted in the figure yield a
significant reduction in the value of neutron skin of ${}^{208}$Pb:
from $R_{n}\!-\!R_{p}\!=\!0.28$~fm to
$R_{n}\!-\!R_{p}\!=\!0.20$~fm. Note that the neutron skin
$R_{n}\!-\!R_{p}$ is defined as the difference between the (point)
root-mean-square neutron and proton radii.

\subsection{Atomic Parity Violation}
\label{APV}

Due to the widespread interest of the PAVI06 audience on atomic parity
violation, we add a brief discussion that the impact of a 1\%
measurement of the neutron radius in $^{208}$Pb could have on the
neutron radius of those heavy nuclei that have been identified as
promising candidates to the atomic parity violation program. These
include Barium, Dysprosium, Ytterbium, and Francium. For more in
depth discussions on atomic parity violation, see the contributions to
these proceedings by Profs. Derevianko, Lintz, Budker, Tsigutkin,
Gwinner, and Sanguinetti.

In large part the choice of suitable atomic systems is the existence
of very close (nearly degenerate) levels of opposite parity that
enhance significantly the parity violating amplitudes. Unfortunately,
parity-violating matrix elements are contaminated by uncertainties in
both atomic and nuclear structure. A fruitful experimental strategy
for removing the sensitivity to the atomic theory is to measure ratios
of parity violation observables along an isotopic chain. This leaves
nuclear-structure uncertainties, in the form of differences in neutron
radii, as the limiting factor in the search for physics beyond the
standard
model~\cite{Pollock:1992mv,Chen:1993fw,Ramsey-Musolf:1999qk}. All
three elements, Barium, Dysprosium, and Ytterbium, have long chains of
naturally occurring isotopes. While the experimental strategy demands
a precise knowledge of neutron radii along the complete isotopic
chain, we only correlate here (as means of illustration) the neutron
radius of $^{208}$Pb to the neutron radius of a member of the isotopic
chain having a closed neutron shell (or subshell). That is, we focus
exclusively on:
${}^{138}$Ba$(Z\!=\!56; N\!=\!82)$,
${}^{158}$Dy$(Z\!=\!66; N\!=\!92)$, and
${}^{176}$Yb$(Z\!=\!70; N\!=\!106)$. 

\begin{figure}
\vspace{0.50in}
\includegraphics[width=3.25in,angle=0]{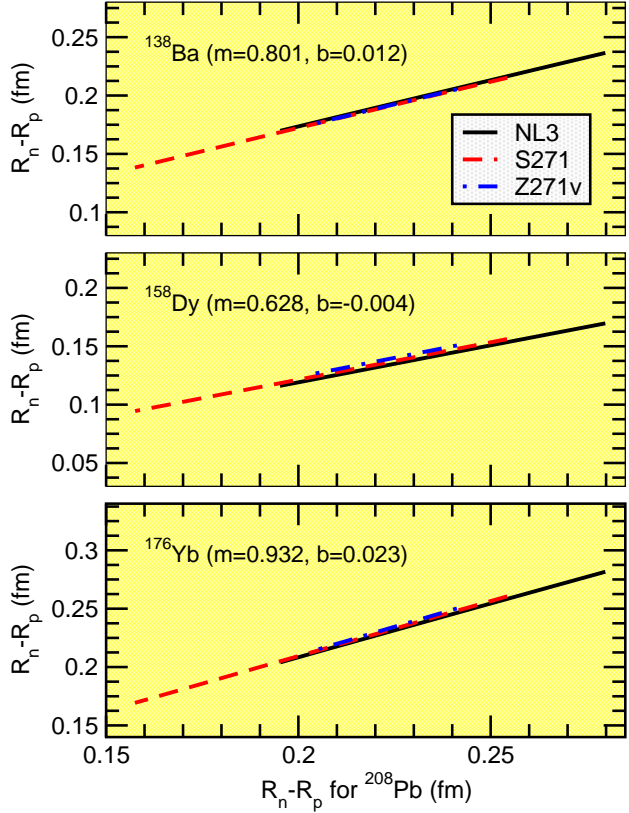}
\caption{Skin-skin correlations for three heavy nuclei of possible relevance 
 to atomic parity violation (${}^{138}$Ba, ${}^{158}$Dy, and ${}^{176}$Yb) 
 as a function of the neutron skin of $^{208}$Pb for three models predicting 
 a different density dependence for the symmetry energy. Quantities in 
 parenthesis represent linear regression coefficients (slope and intercept).}
\label{Fig2}       
\end{figure}

The neutron skins of ${}^{138}$Ba, ${}^{158}$Dy, and ${}^{176}$Yb, are
correlated to the corresponding neutron skin of $^{208}$Pb in the
three panels of Fig.~\ref{Fig2}. We observe a tight linear correlation
that is largely model independent. The linear regression coefficients
(slope $m$ and intercept $b$) have been enclosed in
parenthesis~\cite{Todd:2003xs}.  A theoretical spread of approximately
$0.2$ to $0.3$~fm in the neutron radius of $^{208}$Pb was estimated in
Refs.~\cite{Furnstahl:2001un,Pollock:1992mv}.  Most of this spread is
driven by difference between relativistic and nonrelativistic models,
which has recently been attributed to the poorly known density
dependence of the symmetry
energy~\cite{Piekarewicz:2002jd,Piekarewicz:2003br}. With the
culmination of the the Parity Radius Experiment at the Jefferson
Laboratory~\cite{Michaels:2005}, the theoretical spread will be
replaced by a genuine experimental error that is five times smaller,
that is, $\Delta R_{n}(^{208}{\rm Pb})\simeq 0.056$~fm. This 1\%
measurement of the neutron radius in $^{208}$Pb translates into a
neutron radius uncertainty of $\Delta R_{n}({}^{138}{\rm Ba})\simeq
0.045$~fm, $\Delta R_{n}({}^{158}{\rm Dy})\simeq 0.034$~fm, and
$\Delta R_{n}({}^{176}{\rm Yb})\simeq 0.052$~fm, respectively.  The
results presented above employ a nuclear-structure model that lacks
both deformation and pairing correlations --- effects that may be
important for the nuclei considered in atomic parity violating
experiments. Thus, the present calculations were limited to the study
of that single member of each isotopic chain having a closed neutron
shell. This shortcoming was overcome in Ref.~\cite{Sil:2005tg} through
the inclusion of both nuclear deformation and pairing
correlations. Further, in anticipation of future experiments on high
$Z$ atoms --- where the accuracy in the measurements of atomic parity
violating effects may be significantly improved --- nuclear structure
corrections to the weak charge in Francium isotopes were also
computed. Insofar as the neutron radius of heavy nuclei may be
regarded as a robust bulk property, the results presented in
Ref.~\cite{Sil:2005tg} confirmed the results presented here in
Fig.~\ref{Fig2}.

\subsection{Neutron Star Structure}
\label{neutronstars}

\begin{figure}
\vspace{0.50in}
\includegraphics[width=3.25in,angle=0]{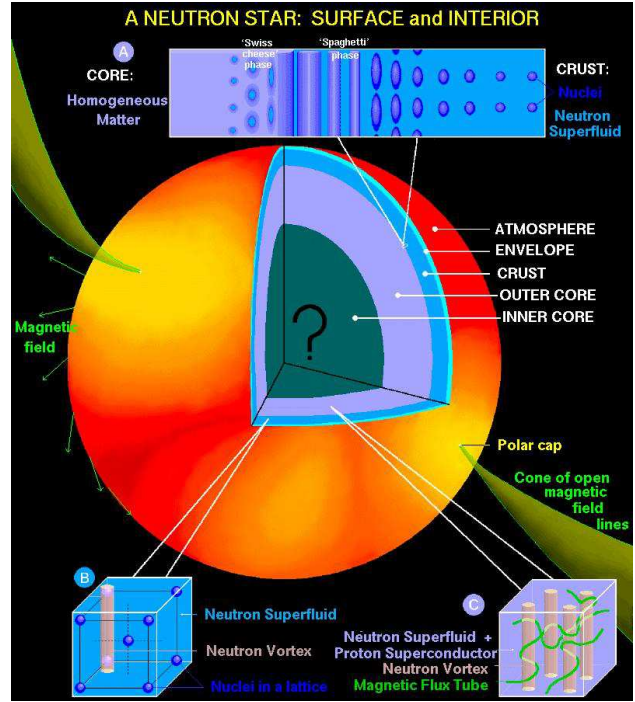}
\caption{State-of-the-art rendition of the structure of a neutron star
         (courtesy of Dany Page).}
\label{Fig3}
\end{figure}

Neutron stars contain a non-uniform crust above a uniform liquid
mantle (or outer core). See Fig.~\ref{Fig3}, courtesy of Dany Page,
for an accurate rendition of the expected structure of a neutron star.
To compute various properties of neutron stars, the equation of state
for the uniform liquid phase is assumed to consist of neutrons,
protons, electrons, and muons in beta equilibrium. Further, it is
assumed that this description remains valid in the high-density inner
core. Thus, the very interesting possibility of transitions to various
exotic phases, such as meson condensates, hyperonic matter, and/or
quark matter, are are not considered here.  In the opposite domain,
namely, at the lower densities of the crust, the uniform system
becomes unstable against density fluctuations. That is, at these
densities it becomes energetically favorable for the system to
separate into regions of high- and low-density matter. In this
non-uniform region the system is speculated to consist of a variety of
complex and exotic structures, such as spheres, cylinders, rods,
plates, {\it etc.} --- collectively dubbed as {\it nuclear
pasta}~\cite{Ravenhall:1983uh,Hashimoto:1984}.  While microscopic
calculations of the nuclear pasta are now becoming
available~\cite{Watanabe:2005qt,Horowitz:2004yf,Horowitz:2004pv,Horowitz:2005zb},
it is premature to incorporate them in our calculation. Hence, after
determining the transition density from the uniform liquid mantle to
the non-uniform solid crust via an RPA stability
analysis~\cite{Horowitz:2000xj}, a simple polytropic equation of state
is used to interpolate between the outer crust~\cite{Baym:1971pw} and
the uniform liquid~\cite{Carriere:2002bx}.

\begin{figure}
\vspace{0.50in}
\includegraphics[width=3.25in,angle=0]{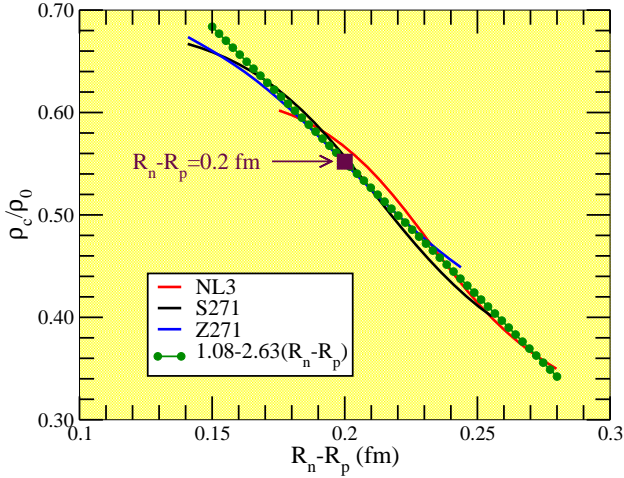}
\caption{Transition density from the uniform liquid mantle to the 
 non-uniform solid crust as a function of the neutron skin of 
 ${}^{208}$Pb. Different models are used to show the largely model 
 independent relation between these two observables.}
\label{Fig4}       
\end{figure}

Results for the transition density from the uniform liquid mantle to
the non-uniform solid crust as a function of the neutron skin in
${}^{208}$Pb are displayed in Fig~\ref{Fig4}. Various models are used
to show the nearly model-independent relation between these two
seemingly distinct observables.  The figure displays an inverse
correlation between the neutron-skin and the transition density found
in Ref~\cite{Horowitz:2000xj}. This correlation suggests that models
with a stiff equation of state predict a low transition density, as it
becomes energetically unfavorable to separate nuclear matter into
regions of high and low densities. Finally, this ``data-to-data''
relation illustrates how an accurate and model-independent
determination of the neutron skin of ${}^{208}$Pb at the Jefferson
Laboratory --- assumed here purely on theoretical biases to be
$R_{n}\!-\!R_{p}\!=\!0.20$~fm --- would determine an important
neutron-star observable.

Having constructed several accurately calibrated equation of states,
such as NL3~\cite{Lalazissis:1996rd} --- together with its softer
versions~\cite{Horowitz:2000xj} --- and
FSUGold~\cite{Todd-Rutel:2005fa}, we now examine their predictions for
the structure of a ``canonical'' 1.4 solar-mass neutron star
($M\!=\!1.4~M_{\odot}$). To do so, we solve the
Tolman-Oppenheimer-Volkoff (TOV) equations, a set of equations
appropriate for the structure of spherically-symmetric neutron stars
in hydrostatic equilibrium. These equations are a generalization of
Newton's equation for hydrostatic equilibrium supplemented by three
corrections from general relativity~\cite{Weinberg:1972}.
Incorporating these correction terms is critical, as typical escape
velocities from neutron stars are of the order of half the speed of
light.  The TOV equations with their associated boundary conditions
are still incomplete without the provision of an equation of state,
{\it i.e.,} a relation between the pressure and the energy density. 
Indeed, if general relativity is assumed valid --- a very modest and 
safe assumption --- the only physics that the structure of neutron 
stars is sensitive to is the equation of state of neutron-rich matter 
in beta equilibrium.

\begin{figure}
\vspace{0.50in}
\includegraphics[width=3.00in,angle=0]{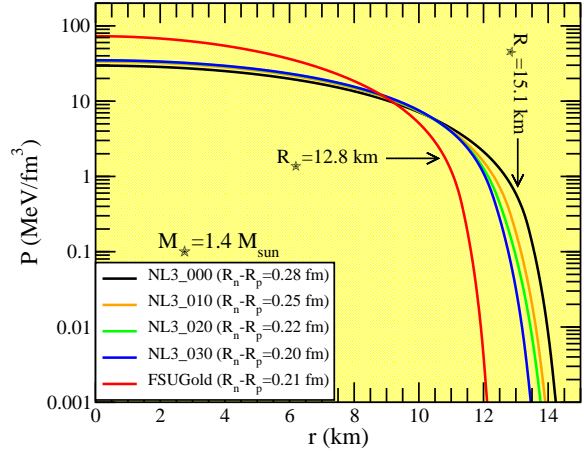}
\caption{Pressure profile of a 1.4 solar-mass neutron star for a variety
 of models predicting a different density dependence of the equation of
 state of neutron-rich matter.}
\label{Fig5}       
\end{figure}

In Fig.~\ref{Fig5} the pressure profile of a $M\!=\!1.4~M_{\odot}$
neutron star is displayed for various equations of states. The curves
labeled as NL3 employ the original NL3 set of Lalazissis and
collaborators~\cite{Lalazissis:1996rd} (denoted here as NL3\_000) plus
variants of it labeled by its value for the isoscalar-isovector mixing
$\Lambda_{\rm v}$ [see Eq.~(\ref{Lagrangian})]. The addition of
$\Lambda_{\rm v}$ allows one to tune the neutron radius of
${}^{208}$Pb without sacrificing the success of the model in
reproducing a variety of well-known ground-state observables, such as
binding energies and charge radii. The figure clearly indicates the
robust correlation between the neutron skin of ${}^{208}$Pb (enclosed
in parenthesis) and the radius of the neutron star: models with a
softer symmetry energy yield both neutron skins and neutron-star radii
smaller relative to the stiffer NL3 set. Yet the correlation, as
opposed to the one displayed in Fig.~\ref{Fig4}, is not model
independent. The radius of the neutron star depends on both the
low- and high-density dependence of the equation of state, while the
neutron skin of ${}^{208}$Pb depends only on the former. Indeed, the
FSUGold set, with a considerably softer high-density component of the
equation of state, requires higher pressures --- and thus higher
densities --- to prevent the gravitational collapse of the star. As a
result, FSUGold predicts a star radius that is considerably smaller
than that of the NL3\_030 set, even though they predict very similar
values for the neutron skin of ${}^{208}$Pb. Many other neutron-star
properties have been correlated to the neutron skin of ${}^{208}$Pb,
such as the cooling of neutron stars through the direct URCA process. 
Unfortunately, due to space limitations they are not being addressed 
here. The interested reader is referred to Ref.~\cite{Todd-Rutel:2005fa} 
for further details.

\section{Conclusions}
\label{conclusions}

The neutron radius of a heavy nucleus is a fundamental
nuclear-structure observable that remains elusive, mainly due to our
inability to use electro-weak probes to sample the neutron
distribution. While a highly mature hadronic program has been used to
map the neutron distribution, the clean extraction of the neutron
radius has been marred by controversial uncertainties in the reaction
mechanism. The established and successful parity-violating program at
the Jefferson Laboratory provides an attractive electro-weak
alternative to the hadronic program. The Parity Radius Experiment at
the Jefferson Laboratory will take advantage of the strong coupling of
the $Z^0$ boson to neutrons to measure the neutron radius of
$^{208}$Pb accurately and model independently. While the intrinsic 
achievement of this experiment is undeniable, in
this contribution we have examined the far-reaching consequences that
such a measurement could have over fields as diverse as atomic parity
violation and astrophysics. First, a tight correlation was found
between the neutron skin of ${}^{208}{\rm Pb}$ (the aim of the PREX
experiment) and the neutron radius of a variety of elements (Barium,
Dysprosium, and Ytterbium) of relevance to atomic parity violation
program. Although the calculations presented here neglect both
deformation and pairing correlations, we have argued --- based on more
sophisticated calculations --- that being a bulk nuclear property, the
neutron radius remains largely insensitive to these effects.  Second,
we demonstrated that PREX will also have a strong impact on various
astrophysical observables. Indeed, a model-independent (or
``data-to-data'') relation between the neutron skin of $^{208}$Pb and
the transition density from the uniform liquid mantle to the
non-uniform solid crust was established. This correlation emerged as a
result of the similar composition of the neutron skin of a heavy
nucleus and the crust of a neutron star: neutron-rich matter at
similar densities. Further, we showed how the measurement of the
neutron skin in $^{208}$Pb is strongly correlated to the radius of a
$M\!=\!1.4~M_{\odot}$ neutron star. This result emerges as a direct
consequence of the density dependence of the symmetry energy, as the
same pressure that pushes neutrons out against surface tension in the
nucleus of $^{208}$Pb is responsible for pushing neutrons out against
gravity in a neutron star. Yet in contrast to the ``data-to-data''
relation described above, this correlation is model dependent --- as
the radius of the neutron star depends on both the low- and
high-density component of the equation of state. Thus, we eagerly
await observational results to constrain the high-density component of
the equation of state. Fortunately, earth- and space-based telescopes
have started to place important constrains on the high-density
component of the equation of state. New telescopes operating at a
variety of wavelengths are turning neutron stars from theoretical
curiosities into powerful diagnostic tools. Significant advances in
observational astronomy will soon yield the first combined measurement 
of mass-radius relations for a variety of neutron stars. These results
--- combined with the Parity Radius Experiment at the Jefferson 
Laboratory --- will provide the most complete information to date on 
the long sought equation of state of neutron-rich matter.

\subsection*{Acknowledgments}
The author wants to acknowledge the many colleagues that have contributed
to this work --- especially Prof. C.J. Horowitz and Dr. Bonnie Todd-Rutel. 
The author is very grateful to the organizers of the PAVI06 conference for 
their kind invitation. This work was supported in part by DOE grant 
DE-FG05-92ER40750.

\bibliographystyle{h-elsevier}

\end{document}